\documentclass{elsart5p}

\usepackage{graphics}
\usepackage{graphicx}
\usepackage{amssymb}

\begin{document}

\begin{frontmatter}



\title{Spiral phases and two-particle bound states from a systematic
  low-energy effective theory for magnons, electrons, and holes in an antiferromagnet}
%

\author[AA]{C. Br\"ugger},
\author[BB]{C.P. Hofmann},
\author[AA]{F. K\"ampfer\corauthref{Name1}},
\ead{fkampfer@itp.unibe.ch}
\author[AA]{M. Moser},
\author[CC]{M. Pepe},
\author[AA]{U.-J. Wiese}

\address[AA]{Institute of Theoretical Physics, Bern University, Bern, Switzerland}
\address[BB]{Facultad de Ciencias, Universidad de Colima, Colima, Mexico}
\address[CC]{Istituto Nazionale di
Fisica Nucleare and Dipartimento di Fisica, Universit\`a di Milano-Bicocca, Milano,
Italy}

\corauth[Name1]{Corresponding and presenting author.}

\begin{abstract}
We have constructed a systematic low-energy effective theory for hole- and electron-doped antiferromagnets, where holes reside in momentum space pockets centered at $(\pm\frac{\pi}{2a},\pm\frac{\pi}{2a})$ and where electrons live in pockets centered at $(\frac{\pi}{a},0)$ or $(0,\frac{\pi}{a})$. The effective theory is used 
to investigate the magnon-mediated binding between two holes or two electrons in an otherwise undoped system. We derive the one-magnon exchange potential from the effective theory and then solve the corresponding two-quasiparticle Schr\"odinger equation. As a result, we find bound state wave functions that resemble $d_{x^2-y^2}$-like or $d_{xy}$-like symmetry. We also study possible ground states of lightly doped antiferromagnets.
\end{abstract}


\end{frontmatter}


\section{Introduction}
\vspace{-1ex}
We have systematically constructed a low-energy effective field theory for hole- and electron-doped antiferromagnets \cite{PRBholes,CondmatElectrons}. The construction relies on basic principles of quantum field theory as well as on a symmetry analysis of the Hubbard or $t$-$J$ model. As an experimental input we take angle resolved photoemission spectroscopy (ARPES) measurements which provide us with the fact that doped holes reside in momentum space pockets centered at $(\pm \frac{\pi}{2a},\pm \frac{\pi}{2a})$ and that electrons live in pockets centered at $(\frac{\pi}{a},0)$ or $(0,\frac{\pi}{a})$. A key ingredient of the effective theory is the nonlinear realization of the spontaneously broken global $SU(2)_s$ symmetry. Using this nonlinearly realized symmetry, the magnons are coupled to the electrons or holes via two vector fields. For reasons of space limitations, here we are not going to show the terms in the effective action. The reader will find all terms properly constructed and listed in \cite{PRBholes} for hole doping and in \cite{CondmatElectrons} for electron doping.

Using the systematic and powerful framework of the low-energy effective theory, we first investigate the magnon-mediated binding of a pair of holes or a pair of electrons in an otherwise undoped antiferromagnet. Solving the corresponding Schr\"odinger equation, we find bound state wave functions that resemble $d$-wave-like symmetry. In a second step we study possible ground states of lightly doped antiferromagnets. In the hole-doped case either a homogeneous phase or a spiral phase is energetically favored. In the electron-doped case the effective theory predicts the absence of a spiral phase.

\section{Hole-Doped Antiferromagnets}
\vspace{-1ex}

Let us now focus on the one-magnon exchange between two holes originating from two different pockets, i.e. one hole from $k^\alpha=\pm(\frac{\pi}{2a}, \frac{\pi}{2a})$ and the other hole from $k^\beta=\pm(\frac{\pi}{2a}, -\frac{\pi}{2a})$. The two holes have antiparallel spin, i.e. spin $+$ and $-$, otherwise no magnon could be exchanged. Evaluating the Feynman diagram for one-magnon exchange and integrating out the magnons, the long-range behavior of the resulting potential is given by
\begin{equation}
V(\vec r \,) =\frac{\Lambda^2}{2\pi\rho_s} \frac{\cos(2\varphi)}{r^2}, 
\end{equation}
where $\vec r=\vec r_+ - \vec r_-$ is the distance vector between the two holes and $\varphi$ is the angle between $\vec r$ and the $x_1$-axis, while $\Lambda$ determines the strength of the hole-one-magnon coupling. Note that the one-magnon-exchange flips the spin of the holes, such that also $\vec r$ is flipped, i.e. 
$\vec r \,'=-\vec r$. We now investigate the corresponding two-component Schr\"odinger equation for the relative motion of the two holes. The two components are the probability amplitudes for the flavor-spin combinations $\alpha_+ \beta_-$ and $\alpha_- \beta_+$. Due to the $1/r^2$-dependence of the potential, the Schr\"odinger equation separates into an angular- and a radial equation, which both have analytical solutions. We find that the ground state is two-fold degenerate. Combining the two ground state wave functions for the two-hole bound state to an eigenstate of 90 degrees rotation leads to the probability distribution shown in Fig. 1 resembling $d_{x^2-y^2}$-like symmetry.
\begin{figure}
\begin{center}
\includegraphics[width=0.43\textwidth]{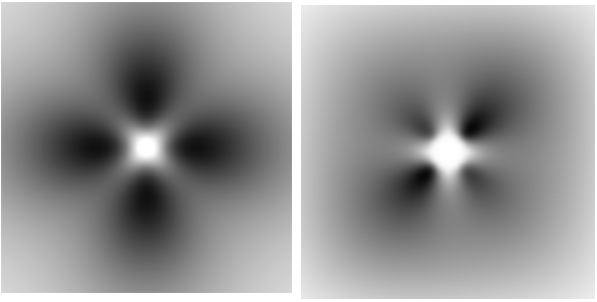}
\end{center}
\caption{Left: Probability distribution for the ground state of two holes. Right: Probability distribution for the ground state of two electrons with total momentum $\vec P =\frac{1}{\sqrt{2}}(P,P)$ } \label{fig1}
\end{figure}

The effective theory is also a very powerful tool to investigate the regime of small doping. In this regime we study possible ground states of the hole-doped system \cite{PRBspirals}. We allow candidate configurations of the staggered magnetization that lead to homogeneous fermion densities. We have proven that this is possible only if the staggered magnetization is constant itself or if it describes a time-independent spiral. In the pure magnon sector, due to the spin stiffness $\rho_s$, a spiral costs more energy than the homogeneous configuration. However, due to the hole-one-magnon vertex in the effective action \cite{PRBholes}, some of the fermions can gain energy in the spiral background, which provides a mechanism for stabilizing a spiral. Based on analytical calculations, in \cite{PRBspirals}, we work out how the hole pockets are populated and whether a homogeneous or a spiral phase in the staggered magnetization is realized depending on the low-energy parameters $\rho_s$, $\Lambda$ and the hole mass. We find that for large $\rho_s/\Lambda^2$ a homogeneous phase is realized, while for intermediate values of $\rho_s/\Lambda^2$ a spiral along a lattice axis is realized. Both configurations of the staggered magnetization are shown in Fig. 2. For small values of $\rho_s/\Lambda^2$ a yet unidentified inhomogeneous phase is energetically favored.

\section{Electron-Doped Antiferromagnets}
\vspace{-1ex}
Using the effective theory, we also calculate the one-magnon exchange potential between two electrons with antiparallel spins. Introducing the total momentum $\vec P$ of the electron pair, the potential is given by
\begin{equation}
V( \vec r \,)  =
\frac{K^2}{2\pi\rho_s}\left[12\frac{\cos(4\varphi)}{r^4}+\frac{P^2}{2}\frac{\cos(2(\varphi+\chi))}{r^2}\right].
\end{equation}
Here $\chi$ is the angle between the total momentum $\vec P$ and the $x_1$-axis, while 
$\vec r = \vec r_+ - \vec r_-$ is the distance vector between the two electrons and $\varphi$ is the angle between $\vec r$ and the $x_1$-axis. The strength of the electron-one-magnon coupling is determined by $K$. Note that for $\vec P=0$, the potential between two electrons is proportional to $1/r^4$, while between two holes it is proportional to $1/r^2$. As a result, for large distances, one-magnon exchange is stronger between two holes than it is between two electrons. Numerically solving the Schr\"odinger equation for a pair of electrons interacting via one-magnon exchange we find the probability distribution of the groundstate, which is shown in Fig. 1 for $\vec P=\frac{1}{\sqrt{2}}(P,P)$. Note that this probability distribution resembles $d_{xy}$-symmetry.

Also in the electron-doped antiferromagnet the effective theory provides a reliable tool to study a lightly doped system. We examine spirals in the staggered magnetization as a possible ground state \cite{CondmatElectrons}. The analysis works similar to the one in the hole-doped case discussed before. However, as a consequence of the weakness of the one-magnon-electron coupling, we find that the electron-doped systems do not favor a spiral in the staggered magnetization.

\begin{figure}
\begin{center}
\includegraphics[width=0.43\textwidth]{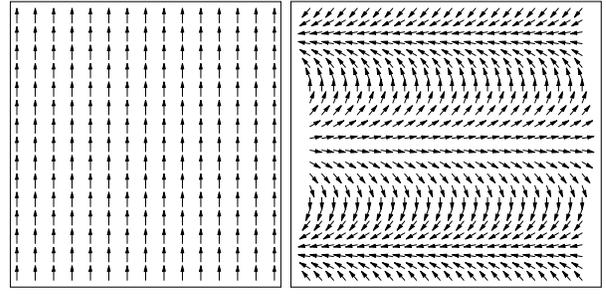}
\end{center}
\caption{Left: Homogeneous phase with constant staggered magnetization; Right: Spiral in the staggered magnetization oriented along a lattice axis.} \label{fig2}
\end{figure}

\section{Summary}
\vspace{-1ex}
We have systematically constructed a low-energy effective theory for magnons and charge carriers in an antiferromagnet. Using the effective theory many interesting aspects of low-energy physics in antiferromagnetic systems can be addressed in a very systematic and transparent manner.

This work was supported in part by funds provided by the Schweizerischer Nationalfonds.


\begin{thebibliography}{99}

\bibitem{PRBholes} C. Br\"ugger, F. K\"ampfer, M. Moser, M. Pepe, and U.-J. Wiese, 
Phys. Rev. B {\bf 74} (2006) 224432.

\bibitem{CondmatElectrons} C. Br\"ugger, C.P. Hofmann, F. K\"ampfer, M. Moser, M. Pepe, and U.-J. Wiese,
Phys. Rev. B {\bf 75}, (2007) 214405.

\bibitem{PRBspirals} C. Br\"ugger, C.P. Hofmann, F. K\"ampfer, M. Pepe, and U.-J. Wiese,
Phys. Rev. B {\bf 75} (2007) 014421.
\end{thebibliography}
\end{document}